\documentstyle[epsfig,rotate,12pt]{article}
\begin{document}
%
%****************************************************************
\def\ss{\footnotesize}
\def\SS{\footnotesize}
\def\sss{\scriptscriptstyle}
\def\barp{{\raise.35ex\hbox{${\sss (}$}}---{\raise.35ex\hbox{${\sss )}$}}}
\def\bdbarp{\hbox{$B_d$\kern-1.4em\raise1.4ex\hbox{\barp}}}
\def\bsbarp{\hbox{$B_s$\kern-1.4em\raise1.4ex\hbox{\barp}}}
\def\dbarp{\hbox{$D$\kern-1.1em\raise1.4ex\hbox{\barp}}}
\def\dcp{D^0_{\sss CP}}
\def\dbar{{\overline{D^0}}}
% Journal and other miscellaneous abbreviations for references
\def \zpc#1#2#3{{\it Z.~Phys.,} C#1 (19#2) #3}
\def \plb#1#2#3{{\it Phys.~Lett.,} B#1 (19#2) #3}
\def \ibj#1#2#3{~#1, (19#2) #3}
\def \prl#1#2#3{{\it Phys.~Rev.~Lett.,} #1 (19#2) #3}
\def \prd#1#2#3{{\it Phys.~Rev.,} D#1 (19#2) #3}
\def \npb#1#2#3{{\it Nucl.~Phys.}, B#1 (19#2) #3}
\def\rly#1{\mathrel{\raise.3ex\hbox{$#1$\kern-.75em\lower1ex\hbox{$\sim$}}}}
\def\lsim{\rly<}
%=======================================================================
\def\bsll{$b \rightarrow s \ell^+ \ell^- $ }
\def\bxsll{$B \rightarrow X_s \ell^+ \ell^- $ }
\def\bxsee{B \rightarrow X_s e^+ e^-  }
\def\bxsmm{B \rightarrow X_s \mu^+ \mu^-  }
\def\bxstt{B \rightarrow X_s \tau^+ \tau^- }
\def\bsee{$b \rightarrow s e^+ e^- $ }
\def\bxsg{$B \rightarrow X_s \gamma $ }
\def\s{\hat{s}}
\def\u{\hat{u}}
\def\z{v \cdot \hat{q}}
\def\dilep{l^+ l^-}
\def\dilpp{l^+ l^+}
\def\dilmm{l^- l^-}
\def\epsm{\epsilon_\mu}
\def\epsn{\epsilon_\nu}
\newcommand{\ra}{\rightarrow}
\newcommand{\bs}{B_s^0}
\newcommand{\bsb}{\overline{B_s^0}}
\def\be{\begin{equation}}
\def\ee{\end{equation}}
\def\g{\gamma}
\newcommand{\BR}{{\cal B}}
\newcommand{\M}{{\cal M}}
\def\mt{m_t}
\def\mb{m_b}
\def\mc{m_c}
% Bra-Kets:
\def\bra{\langle}
\def\ket{\rangle}
\def\bea{\begin{eqnarray}}
\def\eea{\end{eqnarray}}
\def\be{\begin{equation}}
\def\ee{\end{equation}}
% Greek letters:
\def\a{\alpha}
\def\b{\beta}
\def\g{\gamma}
\def\d{\delta}
\def\e{\epsilon}
\def\p{\pi}
\def\ve{\varepsilon}
\def\ep{\varepsilon}
\def\et{\eta}   
\def\l{\lambda} 
\def\m{\mu}
\def\n{\nu}
\def\G{\Gamma}
\def\D{\Delta}
\def\L{\Lambda}  
% Specials:
\def\mt{m_t}
\def\ml{m_\ell}
\def\Mw{M_W}
\def\dBR{{d^2 {\cal B}} \over {d {\hat s} dz}}
\def\Leff{L_{eff}}
\def\bdull{ B_{d,u} \rightarrow  X_s + l^+ l^-}
\def\bduee{ B_{d,u} \rightarrow  X_s + e^+ e^-}
\def\bdumm{ B_{d,u} \rightarrow  X_s + \mu^+ \mu^-}
\newcommand{\bgamaxs}{$B \to X _{s} + \gamma$}
\newcommand{\brogam}{\ $B \to \rho+ \gamma$}
\newcommand{\bdgam}{\ $b \to d+ \gamma$}
\newcommand{\bsg}{\ $b \to s+ g$}
\newcommand{\bsdgam}{\ $b \to (s,d)+ \gamma$}
\newcommand{\bsdell}
   {$b \to (s,d)+ ~\ell \bar{\ell}$ ($\ell = e, \mu, \tau, \nu$)}
\newcommand{\bksell}
   {$B \to K^*+ ~\ell \bar{\ell}$ ($\ell = e, \mu, \tau, \nu$)}
\newcommand{\ope}{operator product expansion}
\newcommand{\bsdnu}
   {$b \to (s,d)+ ~\nu \bar{\nu}$}
\newcommand{\broell}
   {$B \to \rho+ ~\ell \bar{\ell}$ ($\ell = e, \mu, \tau, \nu$)}
\newcommand{\bkell}
  {$B \to K+ ~\ell \bar{\ell}$ ($\ell = e, \mu, \tau, \nu$)}
\newcommand{\bpiell}
 {$B \to \pi+ ~\ell \bar{\ell}$ ($\ell = e, \mu, \tau, \nu$)}
\newcommand{\BGAMAXS}{B \ra X _{s} + \gamma}
\newcommand{\BGAMAXD}{B \ra X _{d} + \gamma}
\newcommand{\BBGAMAXS}{{\cal B}(B \ra  X _{s} + \gamma)}
\newcommand{\BBGAMAXD}{{\cal B}(B \ra  X _{d} + \gamma)}
\newcommand{\BBGAMARHO}{{\cal B}(B \ra  \rho + \gamma)}
\newcommand{\BBGAMAKSTAR}{{\cal B}(B \ra  K^{\star} + \gamma)}
\newcommand{\BGAMARHO}{B \ra  \rho + \gamma}
\newcommand{\BGAMAKSTAR}{B \ra  K^{\star} + \gamma}
\newcommand{\GGAMAXD}{\Gamma(B \ra  X _{d} + \gamma)}
\newcommand{\BGAMAS}{b \ra s + \gamma}
\newcommand{\BGAMAD}{b \ra d + \gamma}
\newcommand{\BBGAMAS}{{\cal B}(b \ra s + \gamma)}
\newcommand{\BBKSTAR}{{\cal B}(B \ra K^\star + \gamma)}
\newcommand{\BKSTAR}{B \ra K^\star + \gamma}
\newcommand{\BBGAMAD}{{\cal B}(b \ra d + \gamma)}
\newcommand{\BGAMAGS}{ b \ra s  + g + \gamma}
\newcommand{\BGAMAGD}{ b \ra d  + g + \gamma}
\newcommand{\GGAMAXS}{\Gamma (B \ra  X _{s} + \gamma)}
\def\Vcd{V_{cd}}
\def\beq{\begin{equation}}
\def\eeq{\end{equation}}
\def\Vcdabs{\vert V_{cd} \vert}
\def\Vus{V_{us}}
\def\Vusabs{\vert V_{us} \vert}
\def\Vcs{V_{cs}}
\def\Vcsabs{\vert V_{cs} \vert}
\def\Vud{V_{ud}}
\def\Vudabs{\vert V_{ud} \vert}
\def\Vbc{V_{cb}}
\def\Vbcabs{\vert V_{cb} \vert}
\def\Vcbabs{\vert V_{cb} \vert}
\def\Vbu{V_{ub}}
\def\Vbuabs{\vert V_{ub}\vert}
\def\Vubabs{\vert V_{ub}\vert}
\def\Vtd{V_{td}}
\def\Vtdabs{\vert V_{td} \vert}
\def\Vts{V_{ts}}
\def\Vtsabs{\vert V_{ts} \vert}
\def\Vtb{V_{tb}}
\def\Vtbabs{\vert V_{tb}\vert}
\newcommand{\bd}{B_d^0}
\newcommand{\bdb}{\overline{B_d^0}}
\newcommand{\abseps}{\vert\epsilon\vert}
\newcommand{\bsdem}
   {$b \to (s,d)+ ~\ell \bar{\ell}$ ($\ell = e, \mu$)}
%%%%%%%%%%%%%%%%%%%%%%%
\def\MAT#1#2#3{<{#1}\vert {#2}\vert {#3}>}
%%%%%%%%%%%%%%%%%%%%%%%Sheldon's defs.
\def\BZ{B_d^0}
\def\BZB{\bar{B_d^0}}
\def\BS{B_s^0}
\def\BSB{\bar{B_s^0}}
\def\BP{B_u^+}
\def\BM{B_u^-}
\def\UPS{\Upsilon}
\def\US{\Upsilon(1\hbox{S})}
\def\USS{\Upsilon(2\hbox{S})}
\def\USSS{\Upsilon(3\hbox{S})}
\def\USSSS{\Upsilon(4\hbox{S})}
\def\pppi{p\bar{p}\pi^-}
\def\pppipi{p\bar{p}\pi^+\pi^-}
\newcommand{\fbb}{f^2_{B_d}B_{B_d}}
\newcommand{\fbbs}{f^2_{B_s}B_{B_s}}
\newcommand{\fbd}{f_{B_d}}
\newcommand{\fbs}{f_{B_s}}
\newcommand{\go}[1]{\gamma^{#1}}
\newcommand{\gu}[1]{\gamma_{#1}}
\newcommand{\xeta}{\eta \hspace{-6pt} / }
\newcommand{\xeps}{\epsilon \hspace{-5pt} / }
\newcommand{\xr}{r \hspace{-5pt} / }
\newcommand{\delmd}{\Delta M_d}
\newcommand{\delms}{\Delta M_s}
\newcommand{\ps}{10^{-12} s}
\def\lp{l^+}
\def\lm{l^-}
\def\sw{\sin{^2}\theta _{W}}
\def\sh{\hat s}
\def\mh{\hat m}
%%%%%%%%%%%%%%%%%%%%%%%spinor
\def\qbar{\overline q}
\def\sLbar{\overline s_L}
\def\ubar{\overline u}
\def\sbar{\overline s}
\def\dbar{\overline d}
\def\bbar{\overline b}
\def\cbar{\overline c}
\def\bLbar{\overline b_L}
\newcommand{\bu}{B_u^\pm}
\def\qq{\qbar{{\lambda_a}\over 2} q}
\def\q5q{\qbar{{\lambda_a}\over 2} i\gamma_5 q}
\def\eLbar{\overline l_L}
\def\eRbar{\overline l_R}
\def\nulbar{\overline {\nu_\ell}}
\def\nul{\nu_\ell}
\def\gmu{\gamma_\mu}
\def\gmuu{\gamma^\mu}
\newcommand{\bsgam}{\ $b \to s+ \gamma$}
\newcommand{\bsdg}{\ $b \to (s,d) + g$}
\newcommand{\bsggam}{\ $b \to s+ \gamma+ g$}
\newcommand{\bdggam}{\ $b \to d+ \gamma+ g$}
\newcommand{\bsdggam}{\ $b \to (s,d)+ \gamma+ g$}
\newcommand{\btod}{$b \to d \gamma, d \gamma g $ }
\newcommand{\btos}{$b \to s \gamma, s \gamma g $ }
\newcommand{\bgamaxd}{$B \to X _{d} + \gamma$}
\newcommand{\brbgamaxd}{{\cal B}(B \to X _{d} + \gamma)}
\newcommand{\bksgam}{\ $B \to K^*+ \gamma$}
%%%%%%%%%%%%%%%%%%%%%%%%%%%%%%%%%%%%%%%%%%%%%%%%%%%%
\def\to{\rightarrow}
\def\VA{(V-A)}
\def\ij{_{ij}}
\def\ji{_{ji}}
\def\dbrp{{dB^+\over d\sh}}
\def\dbrm{\int_0^1 dz {d^2 B \over {d\sh dz}} -
          \int_{-1}^0 dz {d^2 B \over {d\sh dz}} }
\def\ddb{{d^2 B \over d\sh dz}}
\def\mb{m_b}
\def\xt{x_t}
\def\xs{x_s}
\def\xd{x_d}
\newcommand{\kkbar}{$K^0$-${\overline{K^0}}$}
\newcommand{\bdbdbar}{$B_d^0$-${\overline{B_d^0}}$}
\newcommand{\bsbsbar}{$B_s^0$-${\overline{B_s^0}}$}
\def\Cz{{\overline C}^Z}
\def\Cb{{\overline C}^{Box}}
\def\Fs{{F_1}^{(s)}}
\def\Fl{{F_1}^{(l)}}
\def\Fm{{F_2}}
\def\as{\alpha _s}
\def\alfas{\alpha _s}
\def\etal{et al.}
%%%%%%%%%%%%%%%%%%%%%%%%%%%%%%%%%%%%%%%%%%%%%%%%%%%%%%%%%%%%%%%%%%%%%
\def\si{sin\theta_c}
\def\co{cos\theta_c}
\def\BDl{B \to D \ell \nu_\ell}
\def\BDSl{B \to D^* \ell \nu_\ell}
\def\BPIl{B \to \pi \ell \nu_\ell}
\def\BROl{B \to \rho \ell \nu_\ell}
\def\JMUl{J^\mu ^{lept.}}
\def\JMUh{J^\mu ^{hadr.}}
\def\vdvp{v \cdot v^\prime}
\def\xip{\xi_+(\vdvp )}
\def\xim{\xi_-(\vdvp )}
\def\xiv{\xi_V(\vdvp )}
\def\xiao{\xi_{A_1}(\vdvp )}
\def\xiaoo{\xi_{A_1}(\vdvp =1 )}
\def\xiat{\xi_{A_2}(\vdvp )}
\def\xiath{\xi_{A_3}(\vdvp )}
\def\FP{F_+ (q^2)}
\def\FM{F_- (q^2)}
\def\VV{V (q^2)}
\def\Vbc{V_{cb}}
\def\Vbu{V_{ub}}
\def\Vtd{V_{td}}
\def\Vts{V_{ts}}
\def\Vtb{V_{tb}}
\def\pp{{\prime\prime}}
%%% Add here if the photon is to be included in superscripts of F_1:
\def\pg{}
%%%%%%%%%%%%%%%%%%%%%%%%%%%%%%%%
%
\begin{titlepage}

\large
\centerline {\bf Rare $B$ Decays in the Standard Model }
\normalsize

\vskip 2.0cm
\centerline {A.~Ali~\footnote{ali@x4u2.desy.de}}
\centerline { Deutsches~Elektronen-Synchrotron~DESY,
     Notkestrasse~85,~D-22603~Hamburg,~FRG}
\vskip 4.0cm

\centerline {\bf Abstract}
\vskip 1.0cm
We discuss the electromagnetic-penguin-dominated radiative $B$ decays
 $B \to X_s + \gamma, ~B^{\pm (0)} \to K^{*\pm (0)} + \gamma$, and
$B_s \to \phi + \gamma$
 in the context of the standard model (SM) and their 
Cabibbo-Kobayashi-Maskawa (CKM)-suppressed counterparts,
$B \to X_d + \gamma$, $B^\pm \to \rho^\pm + \gamma,
~B^{0} \to (\rho^{0}, \omega) + \gamma$, and $B_s \to K^{* 0} + \gamma$,
 using QCD
sum rules for the exclusive decays. The importance of these decays in 
determining the parameters of the CKM matrix \cite{CKM} is
emphasized. The semileptonic decays $B \to X_s \ell^+ \ell^-$
are also discussed in the context of the SM and their role in determining 
the Wilson coefficients of the effective theory is stressed. Comparison
with the existing measurements are made and SM-based predictions 
for a large number of rare $B$ decays are presented.
\thispagestyle{empty}
\vfill
\end{titlepage}
%\textheight 23.0 true cm
%
%
%%%%%%%%%%%%%%Rare Decays %%%%%%%%%%%%%%%%%%%%%%%%%%%%%%%%%%%%%%%%%%%%%%%%%
%
\newpage
 \section{Estimates of $\BBGAMAXS$ and $\Vtsabs$ in the Standard Model}
\par
  The Standard Model (SM) of particle physics does not admit 
Flavour-changing-neutral-current 
(FCNC) transitions in the Born approximation.
 However, they are induced through the exchange of $W^\pm$ bosons
in loop diagrams. The short-distance contribution in rare decays is 
dominated by the
(virtual) top quark contribution. Hence the decay characteristics provide
quantitative information on the top quark mass and the 
Cabibbo-Kobayashi-Maskawa (CKM) matrix elements $V_{ti}; ~i=d,s,b$ 
\cite{CKM}. We shall discuss  representative examples
from several such transitions involving $B$ decays,
starting with the decay $\BGAMAXS$, which has been measured by CLEO
\cite{CLEOrare2}. This was preceded by the measurement of the exclusive 
decay $\BGAMAKSTAR$ by the same collaboration \cite{CLEOrare1}. The present 
measurements give \cite{CLEOwarsaw}:
\begin{eqnarray}
{\cal B}(\BGAMAXS) &=& (2.32\pm 0.57\pm 0.35)\times 10^{-4} ~,\\
{\cal B}(\BGAMAKSTAR) &=& (4.2\pm 0.8 \pm 0.6)\times 10^{-5}~,
\end{eqnarray}
yielding an exclusive-to-inclusive ratio:
\begin{equation}
R_{K^*} = \frac{\Gamma(\BGAMAKSTAR)}{\Gamma(\BGAMAXS)}=(18.1\pm 6.8)\% ~.
\end{equation}
These decay rates determine the ratio of the CKM matrix 
elements $\Vtsabs/\Vcbabs$ and the quantity $R_{K^*}$ provides information
on the decay form factor in $\BGAMAKSTAR$. In what follows we take up
these points briefly.

The leading contribution to $b \to s +\gamma$ arises
at one-loop from the so-called penguin diagrams. With the help of the
unitarity of the CKM matrix,
the decay matrix element in the lowest order can be written as:
\begin{equation} \label{e2}
 {\cal M }(b \to s ~+\gamma)
    = \frac{G_F}{\sqrt{2}}\,\frac{e}{2 \pi^2} \,\lambda_{t}
   \,(F_2 (x_t)-F_2(x_c))\, q^\mu \epsilon^\nu \bar{s} \sigma_{\mu \nu}
      (m_bR ~+ ~m_sL)b ~.
 \end{equation}
where $x_i= ~m_i^2/m_W^2$, and
$q_\mu$  and $\epsilon_\mu$ are, respectively, the photon four-momentum
and polarization vector.
The GIM mechanism \cite{GIM} is manifest in this amplitude and the
CKM-matrix element dependence is factorized in $\lambda_t\equiv V_{tb} 
V_{ts}^*$.
The (modified) Inami-Lim function $F_2(x_i)$ derived from the (1-loop) 
penguin diagrams is given by \cite{InamiLim}:
\begin{equation}
F_{2}(x) = \frac{x}{24 (x-1)^{4}} \ \left[6 x (3 x -2 )
\log x - (x-1) (8 x^{2} +5 x -7 ) \right]. \nonumber \\
\end{equation}
 The measurement of the branching ratio for $\BGAMAXS$ can be 
readily interpreted in terms of the CKM-matrix element product
$\lambda_t/\Vcbabs$ or equivalently $\Vtsabs/\Vcbabs$.
For a quantitative determination
of $\Vtsabs/\Vcbabs$, however,  QCD radiative
corrections have to be computed and
the contribution  of the so-called long-distance effects estimated.

  The appropriate framework to incorporate
QCD corrections is that of an effective theory obtained by integrating 
out the
heavy degrees of freedom, which in the present context are the top quark 
and $W^\pm$ bosons.
% This effective theory
%is an expansion in $1/m_W^2$ and  involves a tower of
%increasing higher dimensional operators
%built from the  quark fields $(u,d,s,c,b)$, photon, gluons and leptons. The 
%presence
%of the top quark and of the $W^\pm$ bosons is reflected through the
%effective coefficients of these operators which become functions of
%their masses.
 The operator basis depends on the underlying theory and 
for the SM one has (keeping operators up to dimension 6), 
\begin{equation}\label{heffbsg}
{\cal H}_{eff}(b \to s +\gamma) = - \frac{4 G_F}{\sqrt{2}} V_{ts}^* V_{tb}
        \sum_{i=1}^{8} C_i (\mu) {\cal O}_i (\mu) ,
\end{equation}
where the operator basis, the ``matching conditions" $C_{i}(m_W)$, and
the solutions of the renormalization group equations $C_{i}(\mu)$ can be
seen in ref.~\cite{ALI96}.
The perturbative QCD corrections to the decay rate $\GGAMAXS$ have two 
distinct contributions:
\begin{itemize}
\item Corrections to the Wilson coefficients
$C_i(\mu)$, calculated with the help of the 
 renormalization group equation, whose solution requires the
knowledge of the anomalous dimension matrix in a given order in $\as$.
\item Corrections to the matrix elements of the operators
${\cal O}_i$ entering through the effective Hamiltonian
at the scale $\mu=O(m_b)$.
\end{itemize}
The anomalous dimension matrix is needed in order 
to sum up large logarithms, 
i.e., terms 
like $\as^{n}(m_W)\log^{m}(m_b/M)$, where $M=m_t$ or $ m_W$ and $m\leq n$
(with $n=0,1,2,...)$. At present only the leading logarithmic corrections 
$(m=n)$ have been calculated systematically
and checked by several independent groups in the complete basis given 
in Eq.~(\ref{heffbsg})  \cite{Ciuchini}.
First calculations of the NLO corrections to the anomalous dimension matrix
have been recently reported by Misiak \cite{Misiak96} and are found to be 
small. Next-to-leading 
order corrections to the matrix elements are now available completely. 
They are of two kinds: 
 \begin{itemize}
 \item QCD Bremsstrahlung corrections $b \to s \gamma + g$, which are
needed both to cancel the infrared divergences
 in the decay rate for
$\BGAMAXS$ and in obtaining a non-trivial QCD contribution to the
photon energy spectrum in the inclusive decay $\BGAMAXS$.
\item Next-to-leading order virtual corrections to the matrix elements
in the decay $b \to s +\gamma$. 
\end{itemize}
The Bremsstrahlung corrections were calculated in \cite{ag1,ag2}  
in the truncated basis and last year also in the complete 
operator basis \cite{ag95,Pott95}.
The higher order matching conditions, i.e., $C_i(m_W)$, are known up to
the desired accuracy, i.e., up to $O(\as(M_W))$ terms \cite{Yao94}. 
The next-to-leading order virtual corrections have also been calculated 
\cite{GHW96}. They reduce the scale-dependence of the inclusive decay width.
The branching ratio  $\BBGAMAXS$ can be expressed in terms of the
semileptonic decay branching ratio
\begin{equation}
\label{brdef}
{\cal B} ( B \ra  X_{s} \g) = [\frac{\Gamma(B \ra  
\gamma + X_{s})}{\Gamma_{SL}}]^{th}
\, {\cal B} (B \to X\ell \nu_\ell)  \qquad ,
\end{equation}  
where the leading-order QCD corrected expression for $\Gamma_{SL}$
 can be seen in \cite{ALI96}. The leading order $(1/m_b)$ power 
corrections in the heavy quark expansion are identical 
in the inclusive decay rates for  $\BGAMAXS$ and $B \to X \ell \nu_\ell$, 
entering in the
numerator and denominator in the square bracket, respectively, and hence 
drop out.

In Ref.~\cite{ALI96}, the present theoretical
errors on the branching ratio ${\cal B} ( B \ra  X_{s} \g)$ are discussed,
 yielding:
\begin{equation}\label{smbsgbr}
{\cal B} (\BGAMAXS )= (3.20 \pm 0.30 \pm 0.38 \pm 0.32) \times 10^{-4}
\end{equation}
where the first error comes from the combined effect of $\Delta m_t$ and
$\Delta \mu$ (the scale dependence), the second error arises from the
 extrinsic sources
(such as $\Delta(m_b)$, $\Delta(BR_{SL})$), and the third error is an
estimate $(\pm 10\%)$ of the NLO anomalous dimension piece 
in $C_7^{\mathit{eff}}$, the coefficient of the magnetic moment operator.
Combining the theoretical errors in quadrature gives \cite{ALI96}:
\begin{equation}\label{smbsgbrf}
{\cal B} (\BGAMAXS )= (3.20 \pm 0.58) \times 10^{-4},
\end{equation}
which is compatible with the present measurement
${\cal B} (\BGAMAXS )= (2.32 \pm 0.67) \times 10^{-4}$ \cite{CLEOrare2}.
Expressed in terms of the CKM matrix element ratio, one gets
\begin{equation}\label{vtscb}
\frac{\Vtsabs}{\Vcbabs} = 0.85 \pm 0.12 (\mbox{expt}) \pm 0.10 (\mbox{th}),
\end{equation}
which is within errors consistent with unity, as expected from the
unitarity of the CKM matrix.

\section{Inclusive radiative decays \bgamaxd }
\par
The theoretical interest in studying the 
(CKM-suppressed) inclusive radiative decays
\bgamaxd\ lies in the first place in the 
 possibility of determining the parameters of the CKM
matrix. We shall use  the Wolfenstein parametrization \cite{Wolfenstein},
in which case the matrix is determined in terms of the four parameters
$A, \lambda=\sin \theta_C$, $\rho$ and $\eta$.
The quantity of interest in the
decays $B \to X_d + \gamma$ is the end-point photon energy spectrum,
which has to be measured requiring that
 the hadronic system $X_d$ recoiling against the
photon does not contain strange hadrons to suppress the large-$E_\g$
photons from the decay $\BGAMAXS$. Assuming that this is feasible,
one can determine 
 from the ratio of the decay rates
$\BBGAMAXD/\BBGAMAXS$ the CKM-Wolfenstein parameters $\rho$ and $\eta$.
 This measurement was first proposed in \cite{ag2}, where
the photon energy spectra were also worked out.

\indent
 In close analogy
with the \bgamaxs\ case discussed earlier,
the complete set of dimension-6 operators relevant for
the processes $b \to d \gamma$ and $b \to d \gamma g$ 
can be written as:
\begin{equation}
\label{heffd}
{\cal H}_{eff}(b \to d)=
 - \frac{4 G_{F}}{\sqrt{2}} \, \xi_{t} \, \sum_{j=1}^{8}
C_{j}(\mu) \, \hat{O}_{j}(\mu),\quad
\end{equation}
where $\xi_{j} = V_{jb} \, V_{jd}^{*}$ for $j=t,c,u$. The operators
 $\hat{O}_j, ~j=1,2$, have implicit in them CKM factors. In the
Wolfenstein parametrization \cite{Wolfenstein}, one can express these
factors as :
\begin{equation} 
\xi_u = A \, \lambda^3 \, (\rho - i \eta),
~~~\xi_c = - A \, \lambda^3 ,
~~~\xi_t=-\xi_u - \xi_c.
\end{equation}
We note that all three CKM-angle-dependent quantities
$\xi_j$ are of the
same order of magnitude, $O(\lambda^3)$. This aspect 
can be taken into account by defining the operators $\hat{O}_1$ and 
$\hat{O}_2$ entering in ${\cal H}_{eff}(b \to d)$ as follows \cite{ag2}:
\begin{eqnarray}
\label{basis}
&&\hat{O}_{1} =
 -\frac{\xi_c}{\xi_t}(\bar{c}_{L \beta} \go{\mu} b_{L \alpha})
(\bar{d}_{L \alpha} \gu{\mu} c_{L \beta})
 -\frac{\xi_u}{\xi_t}(\bar{u}_{L \beta} \go{\mu} b_{L \alpha})
(\bar{d}_{L \alpha} \gu{\mu} u_{L \beta}) ,\nonumber \\
&& \hat{O}_{2} =
-\frac{\xi_c}{\xi_t}(\bar{c}_{L \alpha} \go{\mu} b_{L \alpha})
(\bar{d}_{L \beta} \gu{\mu} c_{L \beta}) 
 -\frac{\xi_u}{\xi_t}(\bar{u}_{L \alpha} \go{\mu}
b_{L \alpha}) (\bar{d}_{L \beta} \gu{\mu} u_{L \beta}) ,
\end{eqnarray}
with the rest of the operators $(\hat{O}_j;~j=3...8)$ 
defined like their
counterparts ${O}_j$ in ${\cal H}_{eff}(b \to s)$, with the obvious 
replacement
$s \to d$. With this choice, the matching conditions $C_j(m_W)$
 and the solutions
of the RG equations yielding $C_j(\mu)$ become
identical for the two operator bases $O_j$ and $\hat{O}_j$.
The essential difference between  $\GGAMAXS$ and $\GGAMAXD$ 
lies in the matrix elements of the first two operators $O_1$ and $O_2$
(in ${\cal H}_{eff}(b \to s)$) and $\hat{O}_1$ and $\hat{O}_2$ (in 
${\cal H}_{eff}(b \to d)$).
The branching ratio  $\BBGAMAXD$ in the SM  can be written as:
\begin{equation}
\label{branstruc}
\BBGAMAXD = D_1 \lambda^2 \{
(1-\rho)^2 + \eta^2 -(1-\rho) D_2 - \eta D_3 +D_4  \} , \quad
\end{equation}
where the functions $D_i$ depend on the parameters $\mt,\mb,\mc,\mu$,
as well as the others we discussed in the context of ${\cal B}(\BGAMAXS)$.
These functions were first calculated in \cite{ag2} in the leading 
logarithmic 
approximation. Recently, these estimates have been improved in 
\cite{aag96}, making use of the NLO calculations in \cite{GHW96}.
 To get an estimate of the inclusive branching 
ratio, the CKM parameters $\rho$ and $\eta$ have to be constrained from the
unitarity fits. Present data and theory 
restrict them to lie in the following
range (at 95\% C.L.) \cite{al96}:
\begin{eqnarray}
 0.20 &\leq & \eta \leq 0.52 , \nonumber \\
 -0.35 &\leq & \rho \leq 0.35 ~,
\label{rhoetarange}
\end{eqnarray}
which, on using the current lower bound from LEP on
 the $B_s^{0}$ - $\overline{B_s^{0}}$ mass difference  
 $\delms > 9.2$ (ps)$^{(-1)}$ \cite{Gibbons96},
is  reduced to $ -0.25 \leq \rho \leq 0.35$ using 
$\xi_s =1.1$, where $\xi_s$ is the $SU(3)$-breaking parameter $\xi_s = 
f_{B_s} \hat{B}_{B_s}/f_{B_d} \hat{B}_{B_d}$.
The preferred CKM-fit values are
$(\rho,\eta) = (0.05,0.36)$, for which one gets \cite{aag96}   
\begin{equation}
 \BBGAMAXD = 1.63 \times 10^{-5},
\end{equation}
whereas $\BBGAMAXD =8.0 \times 10^{-6}$ and $2.8 \times 10^{-5}$ for the 
choice $\rho=0.35, ~\eta=0.40$ and $\rho=-\eta=-0.25$, respectively.
In conclusion, we note that  
the functional dependence of $\BBGAMAXD$ on the Wolfenstein parameters   
$(\rho,\eta)$ is mathematically different than that of $\delms$. However,
qualitatively they are very similar. From the experimental point of
view, the situation $\rho <0$ is favourable for both the measurements as
in this case one expects (relatively) smaller values for $\delms$ and 
larger   
values for the branching ratio $\BBGAMAXD$, as compared to the $\rho > 0$ 
case which would yield larger $\delms$ and smaller $\BBGAMAXD$.

\vspace*{3.0ex}
\subsection{${\cal B}(B \to V + \gamma )$ and constraints on the CKM 
parameters}

\par
Exclusive radiative
 $B$ decays $B \to V + \gamma$, with $V=K^*,\rho,\omega$, are also 
potentially
very interesting from the point of view of determining the CKM parameters
\cite{abs93}. The extraction of these parameters would, however,  involve a 
trustworthy 
estimate of the SD- and LD-contributions in the decay amplitudes.
\par
  The SD-contribution in the 
 exclusive decays $(B^\pm, B^{0}) \to (K^{*\pm}, K^{* 0})+ \gamma$,
$(B^\pm, B^{0}) \to (\rho^\pm,\rho^{0}) + \gamma$,
$B^{0} \to \omega + \gamma$  and the
corresponding $B_s$ decays, $B_s \to \phi + \gamma $, and
$B_s \to K^{* 0} + \gamma $,
involve the magnetic moment operator ${\cal O}_7$ and the related one 
obtained by the obvious change $s \to d$, $\hat{O}_7$.
The transition form factors governing the radiative $B$ decays
 $B \to V + \gamma$ can be generically  defined as:
\be
 \langle V,\lambda |\frac{1}{2} \bar \psi \sigma_{\mu\nu} q^\nu b
 |B\rangle  =
     i \epsilon_{\mu\nu\rho\sigma} e^{(\lambda)}_\nu p^\rho_B p^\sigma_V
F_S^{B\rightarrow V}(0).
\label{defF}
\ee
Here $V$ is a vector meson
with the polarization vector $e^{(\lambda)}$,
$V=\rho, \omega, K^*$ or $\phi$;
$B$ is a generic
$B$-meson $B^\pm, B^{0}$ or $B_s$, and $\psi$ stands for the
field of a light $u,d$ or $s$ quark. The vectors $p_B$, $p_V$ and
$q=p_B-p_V$
correspond to the 4-momenta of the initial $B$-meson and the
outgoing vector
meson and photon, respectively. In (\ref{defF}) the QCD
renormalization of the $\bar \psi \sigma_{\mu\nu} q^\nu b$ operator
is implied.
 Keeping only the SD-contribution 
 leads to obvious relations among the exclusive 
decay rates, exemplified here by the decay
rates for $(B_u,B_d) \to \rho + \gamma$ and $(B_u,B_d) \to K^* + \gamma$:
\be
\frac{\Gamma ((B_u^\pm,B_d^{0}) \to (\rho^\pm,\rho^{0}) + \gamma)}
     {\Gamma ((B_u^\pm,B_d^{0}) \to (K^{*\pm},K^{* 0}) + \gamma)} 
  = \frac{\vert \xi_t \vert^2}{\vert\lambda_t \vert ^2}
      \frac{\vert F_S^{B \to \rho }(0)\vert^2}
          {\vert F_S^{B \to K^* }(0)\vert^2} \Phi_{u,d}
  \simeq \kappa_{u,d}\left[\frac{\Vtdabs}{\Vtsabs}\right]^2 \,,
\label{SMKR}
\ee
where $\Phi_{u,d}$ is a phase-space factor which in all cases is close to 1
and
 $\kappa_{i} \equiv [F_S(B_i \to \rho \gamma)/F_S(B_i \to K^* 
\gamma)]^2$.
The transition form factors $F_S$ are model dependent.
Estimates of $F_S$ in the QCD 
sum rule approach in the normalization of Eq. (\ref{defF}) range 
between $F_S(B \to K^* \gamma) = 0.31$ (Narison in \cite{bksnsr}) to
$F_S(B \to K^* \gamma) = 0.37$ (Ball in \cite{bksnsr}), with a typical   
error of $\pm 15\%$, and hence are all consistent with each other.
This, for example, gives $R_{K^*}=0.16 \pm 0.05$,
using the result from \cite{abs93}, which is in good agreement with data.
The ratios of the form factors, i.e. $\kappa_i$,
should therefore be reliably calculable as they depend
essentially only on the SU(3)-breaking effects which have been estimated
\cite{abs93,bksnsr}.

The LD-amplitudes in radiative $B$ decays from the light quark 
intermediate states necessarily involve other CKM matrix elements. 
Hence, 
the simple factorization of the decay rates in terms of the CKM factors
involving $\Vtdabs$ and $\Vtsabs$ no longer holds thereby
invalidating the relation (\ref{SMKR}) given above. 
In the decays $B \to V + \gamma$ they  are
induced by the matrix elements of the
four-Fermion operators $\hat{O}_1$ and $\hat{O}_2$ (likewise $O_1$ and 
$O_2$). Estimates of these contributions
 require non-perturbative methods.
 This problem has been investigated
in \cite{wyler95,ab95} using 
 a technique \cite{BBK89}
which treats the photon emission from the light quarks in a theoretically
consistent and model-independent way. This has been combined
with the light-cone QCD sum rule approach to calculate both the SD and LD
--- parity conserving and parity violating --- amplitudes
in the decays $(B^\pm, B^{0}) \to (\rho^\pm,\rho/\omega) + \gamma$.
To illustrate this, we concentrate on the $B^\pm$ decays,
$B^\pm \to \rho^\pm + \gamma$ and take up the neutral $B$ decays
$B^{0} \to \rho (\omega) + \gamma$ at the end.
The LD-amplitude of the four-Fermion operators $\hat{O}_1$, $\hat{O}_2$
is dominated by  the
 contribution of the weak annihilation
of valence quarks in the $B$ meson and it is color-allowed for the
decays of charged $B^\pm$ mesons.
Using factorization, the LD-amplitude in the decay $B_u \to \rho^\pm + 
\gamma$ can be written in terms of the form factors $F_1^L$ and $F_2^L$,
\begin{eqnarray}\label{Along}
{\cal A}_{long} &=&
-\frac{e\,G_F}{\sqrt{2}} V_{ub}V_{ud}^\ast
\left( C_2+\frac{1}{N_c}C_1\right) m_\rho
\varepsilon^{(\gamma)}_\mu \varepsilon^{(\rho)}_\nu
\nonumber\\&&{}\times
 \Big\{-i\Big[g^{\mu\nu}(q\cdot p)- p^\mu q^\nu\Big] \cdot 2 F_1^{L}(q^2)
  +\epsilon^{\mu\nu\alpha\beta} p_\alpha q_\beta
 \cdot 2 F_2^{L}(q^2)\Big\}\,.
\end{eqnarray}
Again, one has to invoke a model to calculate the form factors. Estimates
 from the light-cone QCD sum rules give
\cite{ab95}:
\begin{equation}\label{result}
 F^L_1/F_S = 0.0125\pm 0.0010\,,\quad F^L_2/F_S = 0.0155\pm 0.0010 ~,
\end{equation}
where the errors correspond to the variation of the
Borel parameter in the QCD sum rules. Including other possible 
uncertainties, 
 one expects an accuracy of the ratios in (\ref{result}) of order 20\%.
 The parity-conserving and parity-violating amplitudes turn out
to be numerically close to each other in the QCD sum rule approach, 
$F_1^L\simeq F^L_2 \equiv F_L$,
hence the ratio of the LD- and the SD- contributions reduces to a number 
\cite{ab95}
 \begin{equation}\label{ratio2p}
{\cal A}_{long}/{\cal A}_{short}=
R_{L/S}^{B^\pm\to\rho^\pm\gamma}
\cdot\frac{V_{ub}V_{ud}^\ast}{V_{tb}V_{td}^\ast} ~.
\end{equation}
Using $C_2=1.10$, $C_1=-0.235$, $C_7^{\mathit{eff}}=-0.306$
from Ref.~\cite{ALI96} (corresponding to the scale $\mu=5$ GeV) gives:
\begin{equation}\label{result2}
R_{L/S}^{B^\pm\to\rho^\pm\gamma} \equiv
 \frac{4 \pi^2 m_\rho(C_2+C_1/N_c)}{m_b C_7^{\mathit{eff}}}
\cdot\frac{F_L^{B^\pm \to \rho^\pm \gamma}}{F_S^{B^\pm \to \rho^\pm 
\gamma}}=-0.30\pm 0.07 ~, \end{equation}
which is not small.
 To get a ball-park estimate of the ratio
${\cal A}_{long}/{\cal A}_{short}$, we take the central value from 
the CKM fits, yielding $\Vubabs/\Vtdabs \simeq 0.33$ \cite{al96},
\begin{equation}
|{\cal A}_{long}/{\cal A}_{short}|^{B^\pm\to\rho^\pm\gamma}
= |R_{L/S}^{B^\pm\to\rho^\pm\gamma}|
\frac{|V_{ub}V_{ud}|}{|V_{td}V_{tb}|} \simeq 10\% ~.
\label{bpmld}
\end{equation}
Thus, the CKM factors suppress the LD-contributions.

The analogous LD-contributions to the neutral $B$ decays
$B^{0}\to\rho\gamma $ and $B^{0}\to\omega\gamma $ are
expected to be much smaller.
 The corresponding form factors for the decays
$B^{0} \to \rho^0(\omega)  \gamma$ are obtained from
the ones for the decay $B^\pm\to\rho^\pm \gamma$ discussed above by the
replacement of the light quark charges
 $e_u\to e_d$, which gives the factor $-1/2$; in addition,
and more importantly, the
LD-contribution to the neutral $B$ decays
is colour-suppressed, which reflects itself
through the replacement of the factor
$a_1$  by $a_2$. This yields for the ratio
\begin{equation}
\frac{R_{L/S}^{B^{0}\to\rho\gamma}}{R_{L/S}^{B^\pm\to\rho^\pm\gamma}}=
\frac{e_d a_2}{e_u a_1} \simeq -0.13 \pm 0.05 ,
\end{equation}
where the numbers are based on using
$a_2/a_1 = 0.27 \pm 0.10$ \cite{BH95}. This would then yield 
$R_{L/S}^{B^{0}\to\rho\gamma} \simeq R_{L/S}^{B^{0}\to\omega\gamma}=0.05$,
which in turn gives
\begin{equation}
 \frac{{\cal A}_{long}^{B^{0}\to\rho\gamma}}{{\cal 
A}_{short}^{B^{0}\to\rho\gamma}}\leq 0.02.
\end{equation}
This, as well as the estimate in eq.~\ref{bpmld}, should be taken only as 
indicative in view of the approximations made in 
\cite{wyler95,ab95}. That the LD-effects remain small
in ${B^{0}\to\rho\gamma}$ has been supported in a recent analysis
based on the soft-scattering of on-shell hadronic decay products
$B^{0} \to \rho^0 \rho^0 \to \rho \gamma$ \cite{DGP96},
though this paper estimates them somewhat higher   
(between $4 -8\%$).

\begin{figure}[htb]
\vskip -2.4truein
\centerline{\epsfysize=7in
{\epsffile{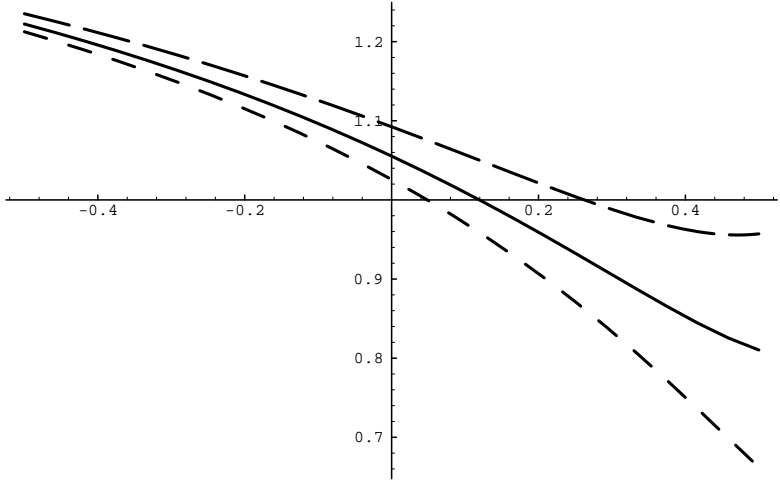}}}
\vskip -1.0truein
\caption[]{
 Ratio of the neutral and charged $B$-decay rates
 $\Gamma (B^\pm \to \rho^\pm \gamma)/2\Gamma (B^{0} \to \rho \gamma)$ as a
function
of the Wolfenstein parameter $\rho$, with $\eta =0.2$ (short-dashed curve),
$\eta =0.3$ (solid curve), and $\eta =0.4$ (long-dashed curve). (Figure taken
from \protect\cite{ab95}.)
\label{abfig2}}
\end{figure}

Restricting to the colour-allowed LD-contributions, the relations, which
obtains ignoring such contributions (and isospin invariance),
\beq\label{ratio2}
\Gamma(B^\pm \to \rho^\pm \gamma)=2 ~\Gamma(B^{0}\to \rho^0  \gamma)
    = 2 ~\Gamma (B^{0} \to \omega  \gamma)~,
\eeq
get modified to
\begin{eqnarray}\label{ratio5}
\lefteqn{\frac{\Gamma(B^\pm\to \rho^\pm\gamma)}{2\Gamma(B^{0}\to \rho\gamma)}
=\frac{\Gamma(B^\pm\to \rho^\pm\gamma)}{2\Gamma(B^{0}\to \omega\gamma)}
 =\left|1+R_{L/S}^{B^\pm\to\rho^\pm\gamma}
\frac{V_{ub}V_{ud}^\ast}{V_{tb}V_{td}^\ast}\right|^2 =
}
\nonumber\\&&{}
=1+2\cdot R_{L/S} V_{ud}\frac{\rho(1-\rho)-\eta^2}{(1-\rho)^2+\eta^2}
+(R_{L/S})^2 V_{ud}^2\frac{\rho^2+\eta^2}{(1-\rho)^2+\eta^2}\,.
\end{eqnarray}
where $R_{L/S}\equiv R_{L/S}^{B^\pm\to\rho^\pm\gamma}$.  
The ratio
$\Gamma(B^\pm\to \rho^\pm\gamma)/2\Gamma(B^{0}\to \rho\gamma)
(=\Gamma(B^\pm\to \rho^\pm\gamma)/2\Gamma(B^{0}\to \omega\gamma))$
is shown in Fig. ~\ref{abfig2}
 as a function of the parameter $\rho$, with
 $\eta= 0.2, ~0.3$  and $0.4$.
This suggests that
a measurement of this ratio would constrain the Wolfenstein parameters
$(\rho, \eta)$, with the dependence on $\rho$ more marked
than on $\eta$. In particular,
a negative value of $\rho$ leads to a
 constructive interference in
$B_u\to\rho\gamma$ decays, while large positive values of $\rho$ give 
a destructive interference.

\par
\begin{figure}[htb]
\vskip -2.4truein
\centerline{\epsfysize=7in
{\epsffile{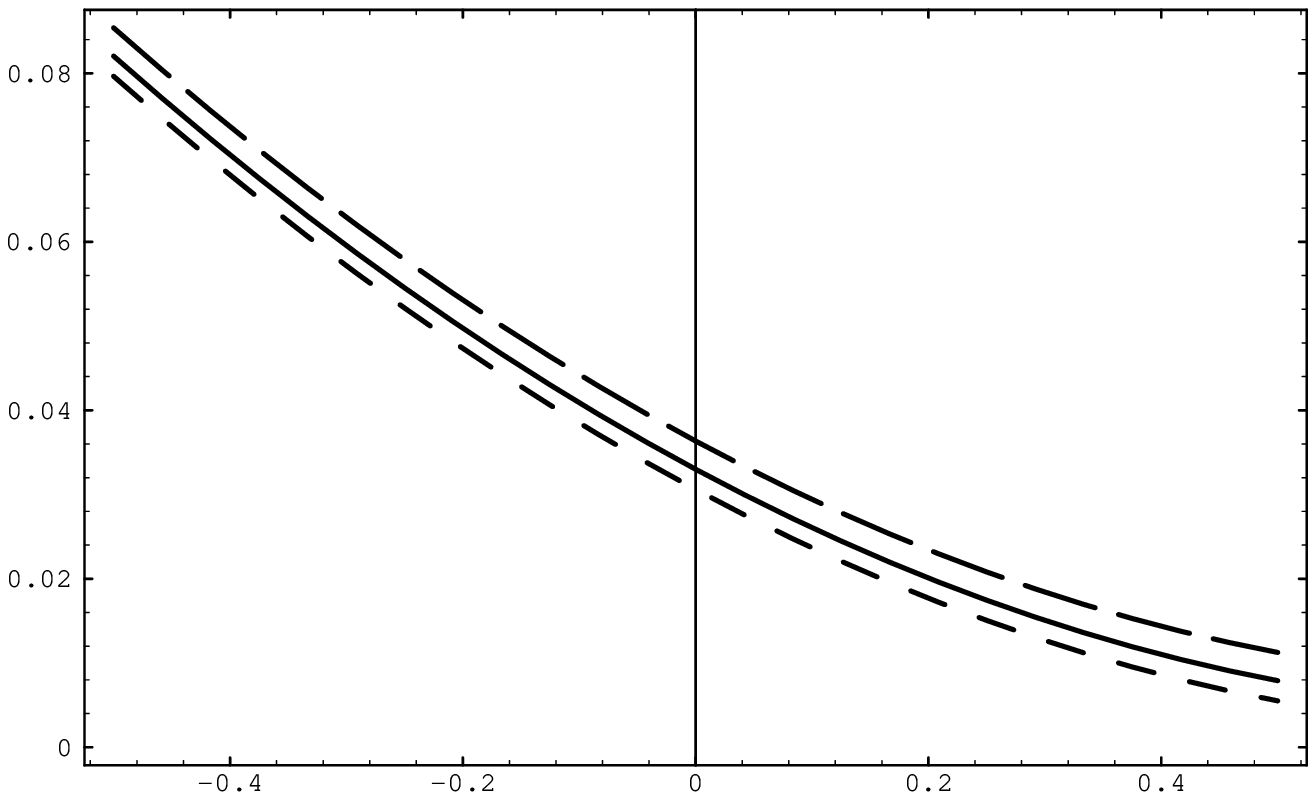}}}
\vskip -1.0truein
\caption[]{
 Ratio of the CKM-suppressed and CKM-allowed radiative $B$-decay
rates
$\Gamma (B_u \to \rho \gamma)/\Gamma (B \to K^* \gamma)$ (with $B=B_u$ or
$B_d$) as a function of the Wolfenstein parameter $\rho$,
a) with $\eta =0.2$ (short-dashed curve), $\eta =0.3$ (solid curve), and
$\eta =0.4$ (long-dashed curve). (Figure taken from \protect\cite{ab95}.)
\label{abfig3}}
\end{figure}

The ratio of the CKM-suppressed and CKM-allowed
decay rates  for charged $B$ mesons
gets modified due to the LD contributions. Following earlier discussion,
we ignore the LD-contributions in $\Gamma(B \to K^*\gamma)$. The ratio of
the decay rates in question can therefore be written as:
\begin{eqnarray}\label{ratio3}
\lefteqn{\frac{\Gamma(B^\pm\to \rho^\pm\gamma)}{\Gamma(B^\pm\to 
K^{*\pm}\gamma)} = \kappa_u \lambda^2[(1-\rho)^2+\eta^2]
}
\nonumber\\&&{}
\times\Bigg\{
1+2\cdot R_{L/S} V_{ud}\frac{\rho(1-\rho)-\eta^2}{(1-\rho)^2+\eta^2}
+(R_{L/S})^2 V_{ud}^2\frac{\rho^2+\eta^2}{(1-\rho)^2+\eta^2}\Bigg\}\,,
\end{eqnarray}
 Using the central value from the estimates of the ratio of the
  form factors squared 
$\kappa_u=0.59 \pm 0.08$
  \cite{abs93}, we show the ratio (\ref{ratio3}) in Fig. 
~\ref{abfig3} as a function of $\rho$ for $\eta=0.2,0.3$, and $0.4$.
It is seen that the dependence of this ratio is rather weak on $\eta$
but it depends on $\rho$ rather sensitively.
The effect of the LD-contributions is modest but not negligible, introducing
an uncertainty  
comparable to the $\sim 15\%$ uncertainty in the overall normalization
due to the $SU(3)$-breaking effects in the quantity $\kappa_u$.

%
% This is Figure 1 (from Ali and Braun)
%
%
% This is Figure 2 (from Ali and Braun)
%
   
\indent
Neutral $B$-meson radiative decays are less-prone to the LD-effects,
 as argued above, and hence one expects that to a good approximation
(say, better than $10\%$) the ratio of the decay rates for neutral $B$ meson 
obtained in the approximation of SD-dominance remains valid \cite{abs93}:
\begin{equation}
\frac{\Gamma(B_d\to \rho\gamma,\omega\gamma)}{\Gamma(B\to K^*\gamma)}
 = \kappa_d\lambda^2 [(1-\rho)^2+\eta^2]~,
\end{equation}
where this relation holds for each of the two decay modes separately.

 Finally, combining the estimates for the LD- and SD-form factors in
\cite{ab95} and
\cite{abs93}, respectively, and restricting the Wolfenstein
parameters in the range $-0.25 \leq \rho \leq 0.35$ and $ 0.2 \leq \eta
\leq 0.4$, as suggested by the CKM-fits \cite{al96}, we give the
following ranges for the absolute branching ratios:
\begin{eqnarray}\label{ratio4}
{\cal B}(B^\pm\to \rho^\pm\gamma)
&=& (1.5 \pm 1.1) \times 10^{-6} ~,
\nonumber\\
{\cal B}(B^{0}\to \rho\gamma) &\simeq& {\cal B}(B^{0}\to \omega \gamma)
= (0.65 \pm 0.35) \times 10^{-6} ~,
\end{eqnarray}
where we have used the experimental value for the branching ratio
${\cal B} (B \to K^* + \gamma)$
\cite{CLEOrare1},
adding the errors in quadrature. The large error reflects the poor
knowledge of the CKM matrix elements and hence experimental determination
of these branching ratios will put rather stringent constraints on the
Wolfenstein parameter $\rho$.

In addition to studying the radiative penguin decays of the $B_u^\pm$ and
$B_d^0$ mesons discussed above, hadron machines such as HERA-B will be in a 
position to study the
corresponding decays of the $B_s^0$ meson and $\Lambda_b$ baryon, such as
$B_s^0 \to \phi + \gamma$ and $\Lambda_b \to \Lambda + \gamma$, which have
not been measured so far. We list below the branching ratios in a number of
interesting decay modes calculated in the QCD sum rule approach in 
\cite{abs93}.
\begin{eqnarray}\label{ratio6}
{\cal B}(B_s\to \phi\gamma)
&=& {\cal B}(B_d\to K^* \gamma)
= (4.2 \pm 2.0) \times 10^{-5} ~,
\nonumber\\
\frac{{\cal B}(B_s\to K^*\gamma)}{{\cal B}(B_d\to K^*\gamma)}
 &\simeq & (0.36 \pm 0.14) \vert \frac{\Vtdabs}{\Vtsabs}\vert^2 \nonumber\\
&\Longrightarrow & {\cal B}(B_s\to K^*\gamma)= (0.75 \pm 0.5) \times 
10^{-6} ~.
\end{eqnarray}

The estimated branching ratios in a number of inclusive and
exclusive radiative $B$ decay modes are given  in Table \ref{tab9},
where we have also listed the branching ratios for $B_s \to \gamma \gamma$
and $B_d \to \gamma \gamma$.

\subsection{Inclusive rare decays $B \to X_s \ell^+ \ell^-$ in the SM}

\par
The decays \bxsll, with $\ell=e,\mu,\tau$, provide a more sensitive search
strategy for finding new physics in rare $B$ decays
than for example the decay \bxsg , which constrains
 the magnitude of $C_7^{\mathit{eff}}$.
 The sign of $C_7^{\mathit{eff}}$, which
 depends on the underlying physics, is not
determined by the measurement of ${\cal B}(\BGAMAXS)$. This sign, which 
in our convention is negative in the SM, is in general model dependent.
It is known (see for example \cite{AGM94}) that
in supersymmetric (SUSY) models, both the negative and positive signs are 
allowed as one scans over the allowed SUSY parameter space.
We recall that for low dilepton masses, the differential decay 
rate for \bxsll ~is dominated by the contribution of the virtual photon 
to the charged lepton pair, which in turn  depends on the
effective Wilson coefficient $C_7^{\mathit{eff}}$.
However, as is well known, the \bxsll ~amplitude in the standard model
has two additional terms, arising from the two FCNC four-Fermi operators,
 \footnote{This also
holds for a large class of models such as MSSM and the two-Higgs doublet
models but not for all SM-extensions. In LR symmetric models, for example, 
there
are additional FCNC four-Fermi operators involved \cite{LRsymmetry}.}
which are not constrained by the $\BGAMAXS$ data.  
Calling their coefficients $C_{9}$ and $C_{10}$, it has been argued in
\cite{AGM94} that the signs and
magnitudes of all three coefficients $C_7^{\mathit{eff}}$, $C_{9}$ and 
$C_{10}$
can, in principle,  be determined from the decays $\BGAMAXS$ and \bxsll .

\par
 The SM-based rates for the decay \bsll , calculated in the free quark decay
approximation, have been known in the LO approximation for some time
\cite{BSGAM}. The LO calculations have the unpleasant
feature that the decay distributions and rates are scheme-dependent.
 The required NLO calculation is in the meanwhile
available, which reduces the scheme-dependence of the LO effects in these
decays \cite{MisiakBM94}. In addition,
long-distance (LD) effects, which are expected to be very important in the
decay \bxsll, have also been estimated from data
 on the assumption that they arise dominantly due to
the charmonium resonances $ J/\psi$ and $\psi'$ through the decay chains
$B \rightarrow X_s J/\psi (\psi',...) \rightarrow X_s \ell^+ \ell^-$.
Likewise, the  leading $(1/{m_b}^2)$ power corrections
to the partonic decay rate and the dilepton invariant mass distribution
have been calculated with the help of the operator product expansion in the 
effective heavy quark theory \cite{falketalbsll}. The results of 
\cite{falketalbsll} have, however, not been confirmed in a recent 
independent calculation
\cite{AHHM96}, which finds that the power corrections in the branching
ratio ${\cal B}(B \to X_s \ell^+ \ell^-)$ are small (typically
$-1.5\%$). The corrections in the dilepton mass spectrum and the FB
asymmetry are also small over a good part of this spectrum. However, the 
end-point dilepton invariant mass spectrum is not calculable in the
heavy quark expansion and will have to be modeled. Non-perturbative 
effects in \bxsll have been estimated using the Fermi motion model in
\cite{Aliqcd}. These effects are found to be small except for the end-point
dilepton mass spectrum where they change the underlying parton model 
distributions significantly and have to be taken into account in
the analysis of data \cite{AHHM96}.
 
The amplitude for \bxsll is calculated in the effective theory
approach, which we have discussed earlier,  by
extending the operator basis of the effective Hamiltonian
introduced in Eq.~(\ref{heffbsg}):
\begin{eqnarray}\label{heffbsll}
& & {\cal H}_{eff}(b \to s + \gamma ; b \to s + \ell^+\ell^- ) \nonumber\\
  &=& {\cal H}_{eff} (b \to s + \gamma) -\frac{4 G_F}{\sqrt{2}} V_{ts}^* 
V_{tb} \left[ C_9 {\cal O}_9 +C_{10}{\cal O}_{10} \right],
\end{eqnarray}
where the two additional operators are:
\begin{eqnarray}
{\cal O}_9 &=& \frac{\alpha}{4 \pi} \bar{s}_\alpha \gamma^{\mu} P_L b_\alpha 
\bar{\ell} \gamma_{\mu} \ell , \nonumber\\
{\cal O}_{10} &=& \frac{\alpha}{4 \pi} \bar{s}_\alpha \gamma^{\mu} P_L 
b_\alpha \bar{\ell} \gamma_{\mu}\gamma_5 \ell ~.
\end{eqnarray}

The analytic expressions for $C_{9}(m_W)$ and $C_{10}(m_W)$ can be seen
in \cite{MisiakBM94} and will not be given here.
 We recall that the
coefficient $C_9$ in LO is scheme-dependent. However, this is compensated
by an additional scheme-dependent part in the
(one loop) matrix element of ${\cal O}_9$. We call the
sum  $C_9^{\mathit{eff}}$, which is scheme-independent and enters in the 
physical decay amplitude given below,
\begin{eqnarray}
\lefteqn{{\cal M}(b \to s +\ell^+\ell^-) =
 \frac{4 G_F}{\sqrt{2}} V_{ts}^* V_{tb}\frac{\alpha}{\pi}} \nonumber\\
&\times &\left[ C_9^{\mathit{eff}}\bar{s} \gamma^{\mu} P_L b \bar{\ell} 
\gamma_{\mu} \ell
 +C_{10}\bar{s} \gamma^{\mu} P_L b \bar{\ell} \gamma_{\mu}\gamma_5 \ell
- 2C_7^{\mathit{eff}} \bar{s} i\sigma_{\mu \nu} 
\frac{q^\nu}{q^2}(m_bP_R+m_sP_L)b
\bar{\ell} \gamma^{\mu} \ell \right],\nonumber\\
&& {}
\end{eqnarray}
with
\begin{equation}
C_9^{\mathit{eff}} (\hat{s}) \equiv C_9\eta({\hat{s}}) + Y(\hat{s}).
\end{equation}
The function $Y(\hat{s})$ is the one-loop matrix element of ${\cal O}_9$
and can be seen in literature \cite{MisiakBM94,ALI96}.
A useful quantity is the  differential FB asymmetry in the c.m.s. of the
dilepton
defined in refs. \cite{amm91}:
\begin{equation}\label{FBasym}
\frac{d {\cal A}(\hat{s})}{d\hat{s}} = \int_0^1 \frac{d{\cal B}}{dz}
                                      -\int_0^{-1} \frac{d{\cal B}}{dz},
\end{equation}
where $z=\cos \theta$, with $\theta$ being the angle between the lepton
$\ell^+$ and the $b$-quark. This can be expressed as:
\begin{eqnarray}
	{{\rm d}{\cal A}(\hat{s}) \over {\rm d}\hat{s}} & = &
		- {\cal B}_{sl} \frac{3 \alpha^2}{4 \pi^2}
                \frac{1}{f(\hat{m}_c)} u^2 (\hat{s})
		C_{10} \left[ \hat{s}{\cal \Re} ( C_9^{\mathit{eff}}(\hat{s})) +
		2 C^{eff}_7 (1 + \hat{m}_s^2) \right] .
	\label{eqn:dasym}
\end{eqnarray}
 The Wilson coefficients
$C^{eff}_7$, $C^{eff}_9$ and $C_{10}$ appearing in the above equation
and the dilepton spectrum (see, for example \cite{AHHM96})
can be determined from data by solving the partial branching ratio
${\cal B}(\Delta \hat{s})$ and partial FB asymmetry
${\cal A}(\Delta \hat{s})$, where $\Delta \hat{s}$ defines an
interval in the dilepton invariant mass \cite{AGM94}.

 There are
other quantities which one can measure in the decays $B \to X_s \ell^+ 
\ell^-$ to disentangle the underlying dynamics.
 We mention here the longitudinal polarization
of the lepton in \bxsll, in particular in $B \to X_s \tau^+ \tau^-$,
proposed by Hewett \cite{Hewettpol}. In a recent paper, Kr\"uger and Sehgal
\cite{KS196} have stressed that complementary information is contained in
the two orthogonal components of polarization ($P_T$, the component in the
decay plane, and $P_N$, the component normal to the decay plane), both of
which are proportional to $m_\ell/m_b$, and therefore significant
for the $\tau^+ \tau^-$ channel. A third quantity, called energy asymmetry,
proposed by Cho, Misiak and Wyler \cite{CMW96}, defined as
\begin{equation}
{\cal A}=\frac{N(E_{\ell^-} > E_{\ell^+}) - N(E_{\ell^+} > E_{\ell^-})}
              {N(E_{\ell^-} > E_{\ell^+}) + N(E_{\ell^+} > E_{\ell^-})}~,
\end{equation}
where $N(E_{\ell^-} > E_{\ell^+})$ denotes the number of lepton pairs
where $\ell^+$ is more energetic than $\ell^-$ in the $B$-rest frame,
is, however, not an independent measure, as
it is directly proportional to the FB asymmetry discussed above. The relation
is \cite{AHHM96}:
\begin{equation}
\int {\cal A}(\hat{s})= {\cal B} \times A~.
\end{equation}
This is easy to notice if one writes the Mandelstam variable
$u(\hat{s})$ in the
dilepton c.m. and the $B$-hadron rest systems. 

\begin{figure}[htb]
\vskip -0.5truein
\centerline{\psfig{figure=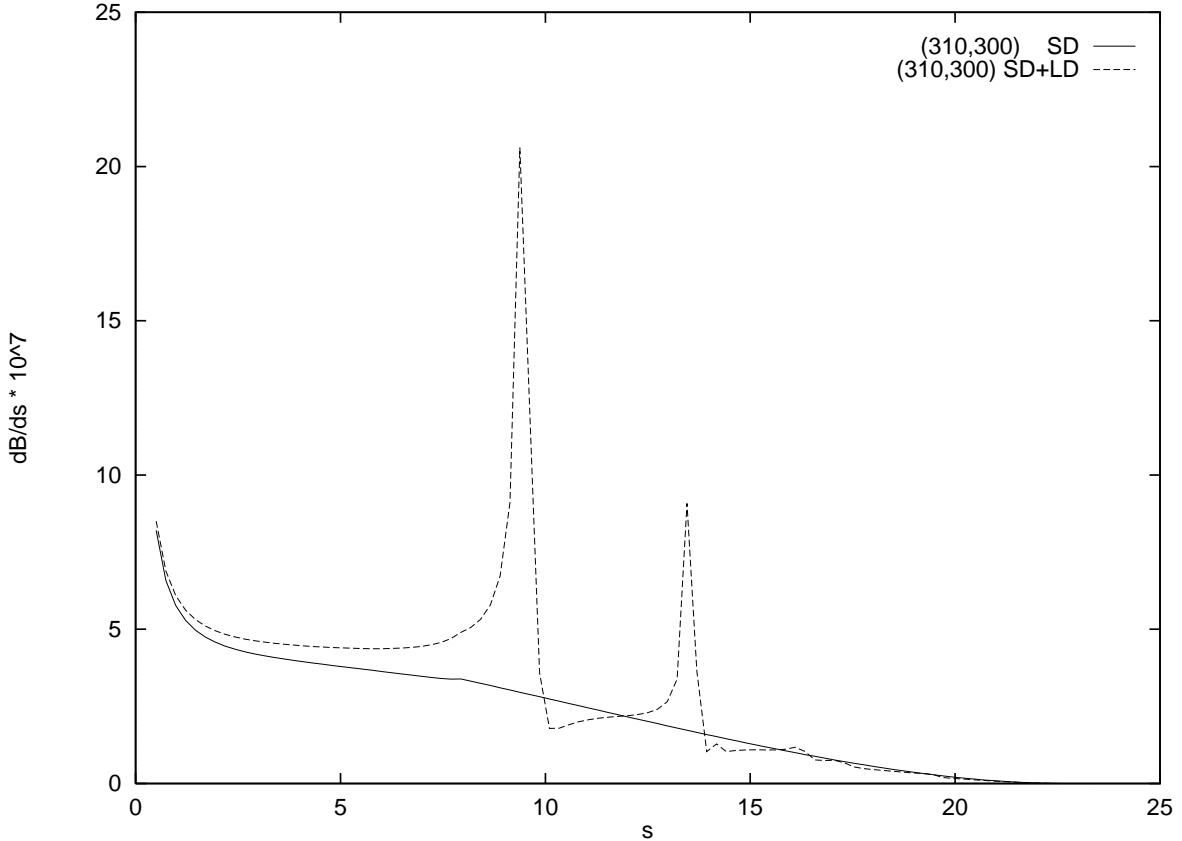,height=16.0cm,angle=270}}
\vskip -0.1truein
\caption[]{
 Dilepton invariant mass distribution in $B \to X_s \ell^+ \ell^-$ in
the SM with the next-to-leading order QCD corrections and non-perturbative
effects calculated in the Fermi motion model (solid curve), and including
the LD-contributions (dashed curve). The model parameters $(p_F,m_q)$
are indicated in the figure. Note that the height of the $J/\psi$ peak
is suppressed due to the linear scale. (Figure taken
 from \protect\cite{AHHM96}.)}
\label{fig:dbrnsm}
\end{figure}

\begin{figure}[htb]
\vskip -0.5truein
\centerline{\psfig{figure=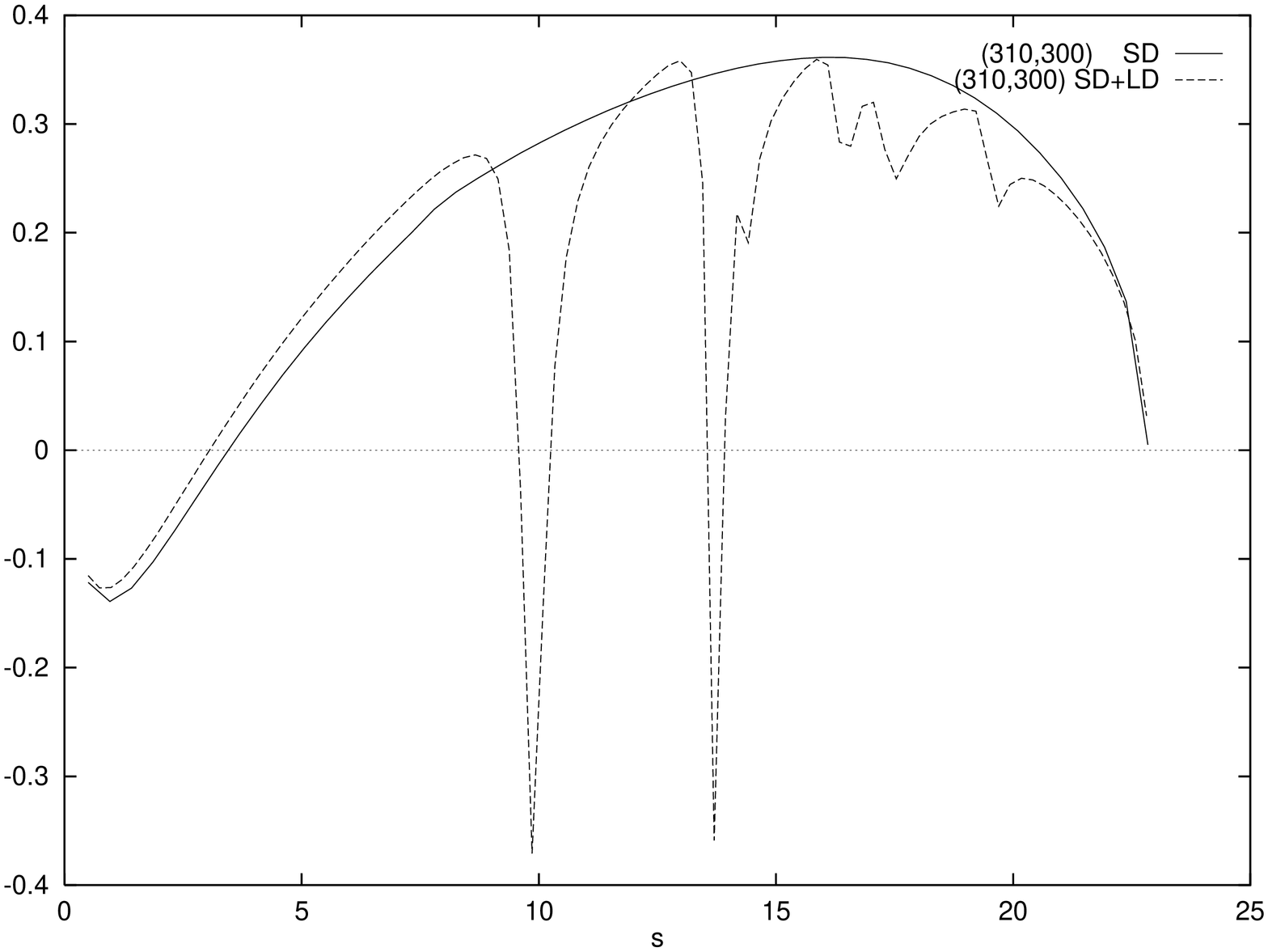,height=16.0cm,angle=270}}
\vskip -0.1truein
\caption[]{
 Normalized FB asymmetry in $B \to X_s \ell^+ \ell^-$ in the SM
 as a function of the dilepton invariant mass calculated using the
 next-to-leading order QCD correction and the Fermi motion effects (solid
curve), and including the LD-contributions (dashed curve). The Fermi motion
parameters are indicated in the figure.
(Figure taken  from \protect\cite{AHHM96}.)}
\label{fig:asymmnsm}
\end{figure}

Next, we discuss the effects of LD contributions in the
processes $B \to X_s \ell^+ \ell^-$. Note that the
 LD contributions due to the vector mesons such as $J/\psi$ and 
$\psi^\prime$, as well as the continuum $c\bar{c}$ contribution already
discussed, 
appear as an effective $(\bar{s}_L \gamma_\mu b_L)(\bar{\ell} \gamma^\mu 
\ell)$ interaction term only, i.e. in the operator ${\cal O}_9$.
 This implies that the LD-contributions should change
$C_9$ effectively,  $C_7$ as discussed earlier is dominated by the
SD-contribution, and 
$C_{10}$ has no LD-contribution. In accordance with this, 
the function $Y(\hat{s})$ is replaced by,
\begin{equation}
	Y(\hat{s}) \rightarrow Y^\prime(\hat{s}) \equiv Y(\hat{s}) + 
		Y_{\mbox{res}}(\hat{s}),
\end{equation}
where $Y_{\mbox{res}}(\hat{s})$ is given as \cite{amm91},
\begin{equation}
	Y_{\mbox{res}}(\hat{s}) = \frac{3}{\alpha^2} \kappa 
		\left(3 C_1 + C_2 + 3 C_3 + C_4 + 3 C_5 + C_6 \right)
		\sum_{V_i = J/\psi, \psi^\prime,...}
		\frac{\pi \Gamma(V_i \rightarrow l^+ l^-) M_{V_i}}{
		M_{V_i}^2 - \hat{s} m_b^2 - i M_{V_i} \Gamma_{V_i}} ,
\end{equation}
where $\kappa$ is a fudge factor, which appears due to the inadequacy
of the factorization framework in describing data on $B \to J/\psi X_s$.
 The long-distance effects lead to significant interference effects
in the  dilepton invariant mass
distribution and the FB asymmetry in \bxsll shown in Figs. \ref{fig:dbrnsm}
and \ref{fig:asymmnsm}, respectively. This can be used to
test the SM, as the signs of the Wilson coefficients in
general are model dependent. For further discussions we refer to 
Ref.~\cite{AHHM96} where also theoretical dispersion on the decay  
distributions due to various input parameters is worked out.
%
% This is Figure 3 (from Ali,Hiller,Handoko and Morozumi)
%
%
%
% This is Figure 4 (from Ali,Hiller,Handoko and Morozumi)
% 
Taking into account the spread in the values of the input parameters,
$\mu, ~\Lambda, ~\mt$, and ${\cal B}_{SL}$
discussed in the previous section in the context of ${\cal B}(B \to X_s + 
\gamma)$, we estimate the following branching ratios for the SD-piece
only (i.e., from the intermediate top quark contribution only) 
\cite{AHHM96}:
 \begin{eqnarray}\label{brbsll}
{\cal B}(\bxsee) &=& (8.4 \pm 2.3) \times 10^{-6}, \nonumber\\
{\cal B}(\bxsmm) &=& (5.7 \pm 1.2) \times 10^{-6}, \nonumber\\
{\cal B}(\bxstt) &=& (2.6 \pm 0.5) \times 10^{-7}, 
\end{eqnarray}
where theoretical errors and the error on ${\cal B}_{SL}$ have been added in 
quadrature.
 The present experimental limit for the inclusive branching ratio in
\bxsll is actually still the one set by the UA1 collaboration some time
ago \cite{UA1R}, namely ${\cal B}(\bxsmm) > 5.0 \times 10^{-5}$. As far
as we know, there are no interesting limits on the other two modes,
involving $X_s e^+e^-$ and $X_s \tau^+ \tau^-$.

  It is obvious from Fig.~\ref{fig:dbrnsm} that   
only in the dilepton mass region far away from
the resonances is there a hope of extracting the Wilson coefficients
governing the short-distance physics. The region below the $J/\psi$ resonance
is well suited for that purpose as the dilepton invariant 
mass distribution there is dominated by the SD-piece.
Including the LD-contributions, following branching ratio has been
estimated for the dilepton mass range $0.2 \leq \hat{s} \leq 0.36$ in 
\cite{AHHM96}:
\be \label{brbslld}
{\cal B}(\bxsmm) = (1.3 \pm 0.3) \times 10^{-6}, 
\ee
with ${\cal B}(\bxsee) \simeq {\cal B}(\bxsmm)$. The FB-asymmetry is 
estimated to be 
in the range $10\%$ - $27\%$, as can be seen in Fig.~\ref{fig:asymmnsm}. 

\par
 The experimental limits on the decay rates of the exclusive decays $B \to 
(K,K^*) \ell^+ \ell^-$ \cite{BH95,Tomasz95}, while arguably closer to the 
SM-based estimates,
can only be interpreted in specific models of form factors, which hinders
somewhat their transcription in terms of the information on
the underlying Wilson coefficients.
 Using the exclusive-to-inclusive ratios
$$R_{K\ell\ell} \equiv \Gamma(B \to K \ell^+ \ell^-)/\Gamma (B \to X_s 
\ell^+ \ell^-) =0.07 \pm 0.02,$$ and
$$R_{K^*\ell\ell} \equiv \Gamma(B \to K^* \ell^+ \ell^-)/\Gamma (B \to X_s
\ell^+ \ell^-) =0.27 \pm 0.0.07,$$ which were estimated in \cite{AGM92}, 
the results are presented in Table \ref{tab9}.
 
 In conclusion, the semileptonic FCNC decays $B \to X_s \ell^+
\ell^-$ (and also the exclusive decays)
 will provide very precise tests of the SM, as they will determine
the signs and magnitudes of the three Wilson coefficients, $C_7,
~C_9^{\mathit{eff}}$ and $C_{10}$.
This, perhaps, may also reveal physics beyond-the-SM
if it is associated with not too high a scale. The MSSM model is a
good case study where measurable deviations from the SM
are anticipated and worked out \cite{AGM94,CMW96}.

%%%%%%%%%%%%%%%%%%%%%%%%%%%%%%%%%%%%%%%%%%%%%%%%%%%%%%%%%%%%%%%%%%%%%%%%%%%
\subsection{Summary and overview of rare $B$ decays in the SM}

\par
 The rare $B$ decay mode $B \to X_s \nu \bar{\nu}$, and some of the
exclusive channels associated with it,
 have comparatively larger branching ratios. The estimated inclusive 
branching ratio in the SM is \cite{AGM92} - \cite{Grossman}:
\begin{equation}\label{bxsnunu}
 {\cal B}(B \to X_s \nu \bar{\nu}) = (4.0 \pm 1.0) \times 10^{-5}~,
\end{equation}
where the main uncertainty in the rates is due
to the top quark mass. The scale-dependence,
which enters indirectly through the top quark mass, has 
been brought under control through the NLL corrections, calculated in
\cite{BuBu93}. The corresponding CKM-suppressed decay $B \to X_d \nu 
\bar{\nu}$ is related by the ratio of the CKM matrix element
squared \cite{AGM92}:
\begin{equation}\label{bxsdnunu}
 \frac{{\cal B}(B \to X_d \nu \bar{\nu})}
  {{\cal B}(B \to X_s \nu \bar{\nu})} = \left[ 
\frac{\Vtdabs}{\Vtsabs}\right]^2 ~.
 \end{equation}
Similar relations hold for the ratios of the exclusive decay rates which
 depend additionally on the ratios of the form factors squared,
which deviate  from unity through $SU(3)$-breaking terms, in close
analogy with the exclusive radiative decays discussed earlier.
 These decays are particularly attractive  probes of
the short-distance physics, as  the long-distance
contributions are practically absent in such decays. Hence, relations
such as the one in (\ref{bxsdnunu}) provide, in principle, one of 
the best methods for the
 determination of the CKM matrix element ratio $\Vtdabs/\Vtsabs$ 
\cite{AGM92}. From the practical point of view, however, these decay 
modes are rather difficult to measure, in particular
at the hadron colliders and probably also at the $B$ factories. The 
best chances are in the $Z^0$-decays at LEP,
from where the present
best upper limit stems \cite{ALEPHwarsaw}:
\begin{equation}\label{bsnunulim}
 {\cal B}(B \to X \nu \bar{\nu}) < 7.7 \times 10^{-4}.
\end{equation}
The estimated branching ratios in a number of inclusive and
exclusive decay modes are given  in Table \ref{tab9}, updating the 
estimates in \cite{ALI96}.

  Further down the entries in Table \ref{tab9} are listed some two-body 
rare decays, such as $(B_s^0, B_d^0) \to \gamma \gamma$, studied in
 \cite{GGTH}, where only the lowest order contributions
are calculated, i.e., without any QCD corrections, and the LD-effects,
which could contribute significantly, are neglected.
The decays $(B_s^0,B_d^0) \to
\ell^+\ell^-$ have been studied in the 
next-to-leading order QCD in \cite{BuBu93}. Some of them,
in particular, the decays $B_s^0 \to \mu^+ \mu^-$ and perhaps also
the radiative decay $B_s^0 \to \gamma \gamma$, have a fighting chance to be
measured at LHC. The estimated decay rates, which depend on the 
pseudoscalar coupling constant $f_{B_s}$ (for $B_s$-decays) and 
$f_{B_d}$ (for $B_d$-decays), together with the present experimental
bounds are listed in Table \ref{tab9}. Since no QCD corrections have been
included in the rate estimates of $(B_{s}, B_{d}) \to \gamma \gamma$,
the branching ratios are rather uncertain.
 The constraints on beyond-the-SM physics that will
eventually follow from these decays are qualitatively similar to the
ones that (would) follow from the decays $\BGAMAXS$ and $B \to X_s \ell^+ 
\ell^-$, which we have discussed at length earlier.

\noindent
\section{Acknowledgements}
 I would like to thank Hrachia Asatrian, Vladimir Braun,
Christoph Greub and Tak Morozumi for helpful 
discussions. The warm hospitality extended  
by Fernando Ferroni and his collaborators in Rome during the 
Beauty 96 workshop is thankfully acknowledged. Finally, I acknowledge the
editorial assistance of Peter Schlein in the preparation of this
manuscript.

%%%%%%%%%%%%%%%%%%%%% REFERENCES %%%%%%%%%%%%%%%%%%%%%%%%%%%%%%%%
\pagebreak

\pagebreak
\newpage

\begin{table}[htb]
\begin{center}
\begin{tabular}{|c|c|c|}  
\hline
Decay Modes & ${\cal B}$(SM) & Measurements and 
         90\% C.L. Upper Limits \\
\hline
$ (B^\pm,B^{0}) \to X_{s} \gamma $
& $(3.2 \pm 0.58)  \times 10^{-4}$ 
& $(2.32 \pm 0.67) \times 10^{-4}$~\cite{CLEOrare2}\\
  \hline
$ (B^\pm,B^{0}) \to K^* \gamma $
& $(4.0 \pm 2.0) \times 10^{-5}$ 
& $(4.2 \pm 1.0) \times 10^{-5}$~\cite{CLEOwarsaw}\\
  \hline
$ (B^\pm,B^{0}) \to X_{d} \gamma $
& $(1.6 \pm 1.2) \times 10^{-5}$ & --\\
  \hline
$ B^\pm  \to \rho^\pm + \gamma $
& $(1.5 \pm 1.1)  \times 10^{-6}$ & $ < 1.1 \times
10^{-5}$~\cite{CLEOwarsaw}\\ \hline
$ B^{0}  \to \rho^0 + \gamma $
& $(0.65 \pm 0.35)  \times 10^{-6}$ & $ < 3.9 \times 10^{-5}$~\cite{CLEOwarsaw}\\ 
\hline
$ B^{0}  \to \omega + \gamma $
& $(0.65 \pm 0.35)  \times 10^{-6}$ & $ <1.3 \times 10^{-5}$~\cite{CLEOwarsaw}\\
\hline
$ B_{s}\to \phi + \gamma $        
                    & $(4.2 \pm 2.0)  \times 10^{-5}$ & --\\
\hline
$ B_{s}\to K^* + \gamma $        
                    & $(0.8 \pm 0.5)  \times 10^{-6}$ & --\\
\hline
$ (B_{d},B_{u}) \to X_{s} e^+ e^- $
                    & $(8.4 \pm 2.2)  \times 10^{-6}$ & --\\
\hline
$ (B_{d},B_{u}) \to X_{d} e^+ e^- $
                    & $(4.9 \pm 2.9) \times 10^{-7}$ & --\\
\hline
$ (B_{d},B_{u}) \to X_{s} \mu^+ \mu^- $
& $(5.7 \pm 1.2)  \times 10^{-6}$ & $  < 3.6 \times 10^{-5}$~\cite{D0warsaw}\\ 
\hline
$ (B_{d},B_{u}) \to X_{d} \mu^+ \mu^- $
& $(3.3 \pm 1.9)  \times 10^{-7}$ &  --\\
\hline
$ (B_{d},B_{u}) \to X_{s} \tau^+ \tau^- $
& $(2.6 \pm 0.5)  \times 10^{-7}$ & --\\
 \hline
$ (B_{d},B_{u}) \to X_{d} \tau^+ \tau^- $
& $(1.5 \pm 0.8)  \times 10^{-8}$ &  --\\
\hline
$ (B_{d},B_{u}) \to K e^+ e^- $
& $(5.9 \pm 2.3)  \times 10^{-7}$ & $ < 1.2 \times 10^{-5}$~\cite{Tomasz95}\\
\hline
$ (B_{d},B_{u}) \to K \mu^+ \mu^- $
 & $(4.0 \pm 1.5)  \times 10^{-7}$ & $ < 0.9 \times 10^{-5}$~\cite{Tomasz95}\\
\hline
$ (B_{d},B_{u}) \to K \mu^+ \mu^- $
 & $(4.0 \pm 1.5)  \times 10^{-7}$ & $ < 0.9 \times 10^{-5}$~\cite{Tomasz95}\\ 
\hline
$ (B_{d},B_{u}) \to K^* e^+ e^- $
                    & $(2.3 \pm 0.9)  \times 10^{-6}$ & $ < 1.6 \times
10^{-5}$~\cite{Tomasz95}\\
\hline
$ (B_{d},B_{u}) \to K^* \mu^+ \mu^- $
& $(1.5 \pm 0.6)\times 10^{-6}$ & $ <2.5 \times 10^{-5}$~\cite{CDF}\\
\hline
$ (B_{d},B_{u}) \to X_{s} ~\nu \bar{\nu} $
& $(4.0 \pm 1.0)  \times 10^{-5}$ & $< 7.7 \times 10^{-4}$~\cite{ALEPHwarsaw}\\
\hline
$ (B_{d},B_{u}) \to X_{d} ~\nu \bar{\nu} $
                    & $(2.3 \pm 1.5)  \times 10^{-6}$ & -- \\
\hline
$ (B_{d},B_{u}) \to K ~\nu \bar{\nu} $
                    & $(3.2 \pm 1.6)  \times 10^{-6}$ & -- \\
\hline
$ (B_{d},B_{u}) \to K^* ~\nu \bar{\nu} $
                    & $(1.1 \pm 0.55)  \times 10^{-5}$ & -- \\
\hline
$ B_{s} \to \gamma \gamma $
    & $(3.0 \pm 1.0)  \times 10^{-7}$ & $ < 1.1 \times 10^{-4}$~\cite{L3}\\ 
\hline
$ B_{d} \to \gamma \gamma $
    & $(1.2 \pm 0.8) \times 10^{-8}$ & $< 3.8 \times 10^{-5}$~\cite{L3}\\
\hline 
$ B_{s} \to \tau^+ \tau^- $
                    & $(7.4 \pm 1.9)  \times 10^{-7}$ & --\\
\hline
$ B_{d} \to \tau^+ \tau^- $
                    & $(3.1 \pm 1.9)  \times 10^{-8}$ & --\\
\hline
$ B_{s} \to \mu^+ \mu^- $
                    & $(3.5 \pm 1.0)  \times 10^{-9}$ & $<8.4 \times
10^{-6}$~\cite{CDF}\\
\hline
$ B_{d} \to \mu^+ \mu^- $
                    & $(1.5 \pm 0.9)  \times 10^{-10}$ & $< 1.6 \times
10^{-6}$~\cite{CDF}\\
\hline
$ B_{s} \to e^+ e^- $
                    & $(8.0 \pm 3.5)  \times 10^{-14}$ & --\\
\hline
$ B_{d} \to e^+ e^- $
                    & $(3.4 \pm 2.3)  \times 10^{-15}$ & --\\
\hline
\end{tabular}
\end{center}
\caption{Estimates of the branching fractions for FCNC $B$-decays
in the standard model taking into account the uncertainties in the
input parameters as discussed in \protect\cite{ALI96}. The entries in the 
second column correspond 
to the short-distance contributions only, except for the radiative decays
$ B^\pm  \to \rho^\pm + \gamma $ and $B^{0} \to (\rho^0, \omega) + \gamma$,
where long-distance effects have also been included.
For the two-body branching 
ratios, we have used $f_{B_d}= 200$ MeV and $f_{B_s}/f_{B_d}=1.16$.
Experimental measurements and upper limits are also listed.
In the second row, the statistical and systematic uncertainties have
been combined to give the quoted experimental uncertainty.}
\label{tab9}
\end{table}

\end{document}